\documentclass[
aip,
rsi,
amsmath,amssymb,
preprint,
]{revtex4-1}

\usepackage{graphicx} 
\usepackage{dcolumn}  
\usepackage{bm}       
\usepackage[dvipsnames,svgnames,x11names]{xcolor}
\usepackage[utf8]{inputenc}
\usepackage[T1]{fontenc}
\usepackage{mathptmx}
\usepackage{setspace} 
\usepackage{placeins}
\usepackage{chemformula}
\setchemformula{kroeger-vink}
\usepackage{textgreek}
\usepackage[markup=underlined]{changes}
\definechangesauthor[color=BrickRed]{AT}
\definechangesauthor[color=NavyBlue]{LW}
\usepackage{xspace}
\usepackage{placeins} 

\usepackage[colorlinks,
            citecolor=blue,
            urlcolor=blue,
            bookmarks=false,
            hypertexnames=true]{hyperref} 
\usepackage[all]{hypcap}

\newcommand{\ta}{\ensuremath{\text{Ta\textsubscript{3}N\textsubscript{5}}}\xspace}
\newcommand{\tuv}{\ensuremath{\text{T-UV-Vis}}\xspace}
\newcommand{\puv}{\ensuremath{\text{P-UV-Vis}}\xspace}

\makeatletter
\renewcommand{\p@subsection}{}
\renewcommand{\p@subsubsection}{}
\makeatother

\begin{document}

\title{Quantifying Thermal, Photovoltage, and Defect Contributions to Transient Absorption of \ta Photoanodes}

\author{Johannes Dittloff}
\affiliation{Walter Schottky Institute, Technical University of Munich, 85748 Garching, Germany}
\affiliation{Physics Department, TUM School of Natural Sciences, Technical University of Munich, 85748 Garching, Germany}

\author{Lukas M. Wolz}
\affiliation{Physics Department, TUM School of Natural Sciences, Technical University of Munich, 85748 Garching, Germany}

\author{Matthias U. Quintern}
\affiliation{Walter Schottky Institute, Technical University of Munich, 85748 Garching, Germany}
\affiliation{Physics Department, TUM School of Natural Sciences, Technical University of Munich, 85748 Garching, Germany}

\author{Laura I. Wagner}
\affiliation{Walter Schottky Institute, Technical University of Munich, 85748 Garching, Germany}
\affiliation{Physics Department, TUM School of Natural Sciences, Technical University of Munich, 85748 Garching, Germany}

\author{Matthias Kuhl}
\affiliation{Physics Department, TUM School of Natural Sciences, Technical University of Munich, 85748 Garching, Germany}

\author{Johanna Eichhorn}
\affiliation{Physics Department, TUM School of Natural Sciences, Technical University of Munich, 85748 Garching, Germany}

\author{Ian D. Sharp$^*$}
\email[sharp@wsi.tum.de]{}
\affiliation{Walter Schottky Institute, Technical University of Munich, 85748 Garching, Germany}
\affiliation{Physics Department, TUM School of Natural Sciences, Technical University of Munich, 85748 Garching, Germany}


\begin{abstract}
\ta is among the most intensively studied photoanode materials for solar-driven water oxidation, yet its performance often remains limited by short carrier lifetimes and defect-mediated recombination. Although transient absorption spectroscopy is widely used to probe carrier dynamics in photoelectrodes, spectral assignments are frequently ambiguous due to overlapping contributions. Here, \textmu s-s transient absorption of \ta thin films is combined with complementary optical spectroscopies to disentangle contributions from lattice heating, electrostatics, and defect states. Photoreflectance reveals three critical points in the \ta band structure, including two anisotropic near-edge transitions at 2.14\,eV and 2.27\,eV and a higher-lying transition near 2.80\,eV, all closely aligned with dominant transient absorption features. A previously unreported photo-induced absorption at 2.80\,eV is attributed to pump-induced lattice heating, while potential-dependent measurements reveal that near-edge bleach features arise from pump-induced band flattening and subsequent surface photovoltage relaxation. Fitting transient absorption spectra with independently measured thermal and electrostatic components enables quantification of both thermal and photovoltage dynamics, while the sub-bandgap response provides insight into the redistribution of defect charge states. Thus, this approach to quantifying thermal, electrostatic, and defect-mediated contributions to \textmu s-s transient absorption provides broadly applicable insights into photoexcitation and relaxation mechanisms in functional semiconductor photoelectrodes.
\end{abstract}

\maketitle 

\section{Introduction}\label{sec:intro}
Solar-driven photoelectrochemical (PEC) water splitting offers a direct route toward sustainable hydrogen production. However, its practical efficiency is often limited by recombination losses and inefficient interfacial charge injection, particularly during the kinetically challenging four-hole oxygen evolution reaction (OER). Addressing these losses requires photoanodes that combine efficient visible light absorption with long-lived hole transport to the semiconductor/electrolyte interface.~\cite{Gratzel.2001,Corby.2021,Jiang.2017,He.2019} In this regard, \ta has emerged as one of the most-intensively studied photoanode materials due to its suitable bandgap of 2.1\,--\,2.2\,eV and favorable band energetics.~\cite{He.2016,Wolz.2024,Higashi.2020,Higashi.2019}
However, \ta photoanodes often fall short of their theoretical solar-to-chemical energy conversion limits due to a combination of inefficient charge transport, short carrier lifetimes, and strong sensitivity to defects and disorder that can introduce deep trap states and facilitate non-radiative recombination losses.~\cite{Eichhorn.2021,Wagner.2024,Wolz.2024,Xiao.2020,Fu.2020} 
To overcome these limitations, a detailed understanding of the charge carrier dynamics is thus critical, particularly at the \textmu s-s timescale where charge accumulation and recombination compete with interfacial charge transfer and photoelectrochemical reaction dynamics.~\cite{Corby.2021,Forster.2020} 

Transient absorption spectroscopy (TA) enables direct observation of photo-induced changes in absorption, offering important insights into carrier populations and defect-mediated recombination dynamics. Conventional interpretations of semiconductor TA spectra invoke pump-induced changes in optical transition probabilities arising from non-equilibrium carrier populations, including band-edge bleach features from state filling and photo-induced absorption from excited-state transitions, as well as electronically active defect states. While such interpretations are physically grounded and have provided important mechanistic insights, TA spectra typically comprise broad and overlapping features whose unambiguous assignment is not straightforward. Indeed, for \ta, previous TA studies have revealed several bleach and photo-induced absorption features in the visible to near-infrared regions, which are commonly assigned to valence band hole populations and transitions associated with nitrogen vacancies and substitutional oxygen states. However, energetic positions, relative amplitudes, and physical assignments of these spectral signatures vary between reports, and can depend strongly on excitation fluence and the probed time window.~\cite{Murthy.2019,Ziani.2015,Yin.2021,Vequizo.2018} As a result, there remains no clear consensus regarding the nature of the excited-state absorption spectrum of \ta, hindering attempts to rationally enhance carrier lifetimes, promote carrier separation, and increase the activity of functional photoelectrodes. 

To address these inconsistencies and establish a rigorous basis for interpreting photo-induced dynamic processes in functional photoelectrodes, we demonstrate that additional physical mechanisms beyond non-equilibrium carrier populations contribute significantly to TA spectra and must be independently characterized for reliable spectral quantification, especially at \textmu s-s time scales. For example, pump-induced lattice heating can modify the optical spectrum through thermally driven redshifting and broadening of interband transitions, potentially generating spectral signatures that can be misassigned to electronic excited-state processes~\cite{Hayes.2016,Miao.2020}. In addition, above-bandgap photoexcitation transiently screens the built-in electric field within the near-surface depletion region, generating a surface photovoltage (SPV) that modulates critical point transition strengths and simultaneously alters near-surface defect charge states. Importantly, both of these effects are particularly significant in the \textmu s-s regime, which represents the natural timescale of interfacial chemical dynamics in photoelectrochemical and photocatalytic systems. However, to date, no systematic effort has been devoted to independently characterize and disentangle these contributions in \ta, resulting in ambiguous excited-state spectral assignments and poor understanding of the nature of its photo-induced dynamics.

Here, we address this challenge by combining \textmu s-s TA of sputtered \ta thin films with three complementary steady-state modulation spectroscopies, each providing independent experimental access to one of the key physical contributions. Starting with photoreflectance (PR) spectroscopy, we first identify the critical point transitions in the electronic structure of \ta, thereby providing reference energies for transient spectral assignments. In addition, we apply potential-dependent UV-Vis (\puv) spectroscopy in a photoelectrochemical cell to isolate the electroabsorption response arising from controlled modulation of the internal electric field within the near-surface depletion region. Finally, we perform temperature-dependent UV-Vis (\tuv) spectroscopy to independently quantify the thermal response as a function of lattice temperature. By fitting TA spectra with these independently measured electrostatic and thermal components, we show that the dominant near- and above-edge transient optical response can be quantitatively reconstructed, enabling simultaneous extraction of the lattice temperature and surface photovoltage dynamics. The remaining sub-gap response is interpreted in terms of non-equilibrium defect charge state redistribution under photoexcitation, providing additional insights into the roles of nitrogen vacancies and substitutional oxygen impurities in the \textmu s-s optical response. Beyond providing a comprehensive understanding of the \textmu s-s TA spectrum of \ta, this systematic approach offers a generalizable strategy for quantifying competing thermal, electrostatic, and defect-mediated contributions to transient optical spectra of semiconductor photoelectrodes, opening new avenues for mechanistic interpretation of photo-induced processes in solar-to-chemical energy conversion systems.

\section{Results and Discussion} \label{sec:results}

\subsection{Structural, photoelectrochemical, and dynamic optical properties}
\ta thin films were prepared by reactive magnetron sputtering followed by high-temperature ammonolysis, as described in the Experimental Section. The 220\,nm thick films were deposited on fused silica with a 7\,nm TiN interlayer, which serves as a semi-transparent conductive back contact that enables electrical biasing during transmission measurements. 
Grazing incidence X-ray diffraction (GI-XRD) confirms the formation of phase-pure orthorhombic \ta (ICDD PDF\# 79-1533), with no detectable secondary phases (Figure~S1~(a)), while atomic force microscopy (AFM) reveals a granular morphology with a root-mean-square (rms) roughness of approximately 2.4\,nm (Figure~S1~(b)), indicating laterally uniform and smooth films. Chopped-illumination linear sweep voltammetry measurements in 1\,M KPi buffer (pH 12.3) with 0.1\,M \ch{K_4[Fe(CN)_6]} as sacrificial hole acceptor exhibit a photoanodic response over the investigated potential range, with a photocurrent density reaching approximately 1.19\,mA\,cm$^{-2}$ at 1.23\,V$_\text{RHE}$ (Figure~S1~(c)), consistent with prior reports for sputtered \ta thin films.~\cite{Wagner.2024,Wolz.2024,Eichhorn.2021} Thus, combined structural and photoelectrochemical characterization verifies the successful fabrication of phase-pure, photo-active \ta, providing a well-defined platform for mechanistic investigation of its photoexcited optical response.

\FloatBarrier
\begin{figure}
    \centering
    \includegraphics[]{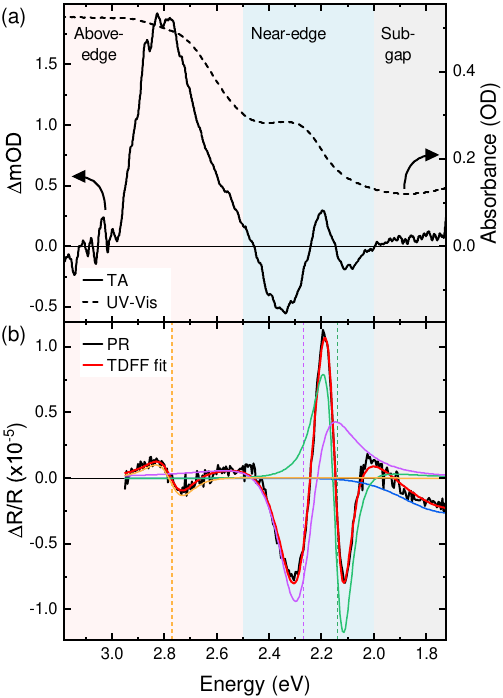} 
    \caption{(a) Ground-state absorption spectrum of the \ta/TiN thin film (right axis) and transient absorption (TA) spectrum recorded at a pump-probe delay of 1\,\textmu s following 355\,nm pulsed excitation with a 1\,mJ\,cm$^{-2}$ pump pulse at 355\,nm (left axis). The transient response exhibits a pronounced photo-induced absorption near 2.80\,eV, two bleach features in the 2.0\,--\,2.5\,eV range, and a broad low-energy photo-induced absorption tail. Colored regions delineate the higher energy above-edge, intermediate energy near-edge, and low energy sub-gap spectral regimes discussed in the text. (b) Photoreflectance spectrum (PR, black) together with a three-component third-derivative functional form (TDFF) lineshape fit including a Gaussian sub-gap tail (red). The individual fitted components (orange, purple, green) are shown along with the extracted critical point energies (vertical lines) and the Gaussian component (blue).}
    \label{fig:res:tas}
\end{figure}

The ground-state optical absorption spectrum of the orthorhombic \ta thin film (\autoref{fig:res:tas}~(a), dashed curve) exhibits an absorption onset near 2.1\,--\,2.2\,eV, consistent with reported values for its fundamental bandgap.~\cite{Wagner.2024,Fu.2020,Xiao.2020} In the near-edge region, the absorbance increases and then transitions to a relatively broad, plateau-like behavior. At higher photon energies, near 2.5\,--\,2.6\,eV, a second increase in absorption is observed, suggesting the onset of additional higher-energy interband transitions. For clarity, we thus refer to three spectral regimes in this work, corresponding to the sub-gap (\textless\,2.0\,eV), near-edge (2.0\,--\,2.5\,eV), and above-edge (\textgreater\,2.5\,eV) regions, as indicated in \autoref{fig:res:tas}~(a).

The corresponding TA spectrum recorded at a pump-probe delay of 1\,\textmu s is shown as the solid curve in \autoref{fig:res:tas}~(a). 
In the above-edge range, a dominant positive feature is observed at approximately 2.80\,eV, with a shoulder extending towards lower energies. In the near-edge region, two bleach features appear at 2.35\,eV and 2.10\,eV, separated by a positive feature at 2.20\,eV. Finally, in the sub-gap region (\textless\,2.0\,eV), a broad but weak photo-induced absorption (PIA) tail extends towards lower energies.

Several of the near-edge and sub-gap features have been reported previously in TA studies of \ta\cite{Yin.2021,Murthy.2019,Vequizo.2018,Ziani.2015}, though their energetic positions, relative amplitudes, and spectral assignments vary between reports. Most prior work has focused primarily on the near-edge region. In particular, the bleach at 2.35\,eV is commonly attributed to the fundamental band-edge transition,~\cite{Murthy.2019,Yin.2021,Ziani.2015} and is frequently interpreted as arising from depletion of the valence band maximum (VBM) due to occupation by photo-generated holes.~\cite{Murthy.2019,Yin.2021} 
In contrast, there remain differing interpretations regarding the nature of the spectral response in the 2.1\,--\,2.2\,eV range. In fs-ps TA measurements, Murthy et al. assigned a bleach near 2.1\,eV to hole trapping into nitrogen vacancy states located just above the VBM, thereby reducing the strength of electronic transitions from these in-gap states to the conduction band minimum (CBM).~\cite{Murthy.2019} 
However, \textmu s-scale TA studies have reported positive absorption features in a similar spectral range that were attributed to hole-related absorption processes.~\cite{Yin.2021,Vequizo.2018} In the same spectral window, Yin et al. also observed a bleach feature that they interpreted as arising from the fundamental bandgap transition along the $a$-axis of the anisotropic \ta structure.~\cite{Yin.2021} These differing interpretations suggest that multiple overlapping contributions may exist in this spectral window, with relative strengths that depend on the measured timescale, excitation density, and sample preparation. Finally, the broad sub-gap PIA tail below 2.0\,eV is typically associated with electronic transitions involving nitrogen vacancy (\ch{v_{N}^{}}) and substitutional oxygen (\ch{O_{N}^{}}) defect states located within the bandgap.~\cite{Murthy.2019,Vequizo.2018,Yin.2021}

Beyond the previously discussed near-edge and sub-gap spectral responses, our measurements reveal a pronounced positive feature in the higher-energy region, centered near 2.80\,eV and extending asymmetrically to lower energies. Complementary photoluminescence spectroscopy (see Figure~S2) shows no pronounced emission in this energy region, consistent with assignment of this feature to PIA. Despite its substantial spectral weight, this contribution has, to the best of our knowledge, not been resolved or analyzed in prior TA studies of \ta, which have largely focused on the visible near-edge region.~\cite{Yin.2021,Murthy.2019,Vequizo.2018,Ziani.2015} Given its position well above the bandgap, this response cannot be straightforwardly explained by conventional band-edge state filling or localized defect-related transitions alone, suggesting that additional excited-state effects must be considered. Taken together, the emergence of this new above-edge response and the differing interpretations for near-edge features reported previously motivate a systematic investigation of the complete excited-state optical response through complementary spectroscopic approaches, as presented below.


\subsection{Identification of critical point transitions}
As a starting point for clarifying the origin of the near-edge and above-edge TA features, we performed PR spectroscopy, which is a well-established technique for identifying critical point transitions in semiconductor band structures. In brief, PR is a contactless modulation spectroscopy that measures optical excitation-induced changes in reflectivity arising from the modulation of internal electric fields. In semiconductor thin films, band bending at surfaces and interfaces gives rise to built-in electric fields that can be transiently modified upon illumination through the establishment of a SPV. The photo-induced modulation of the internal electric field alters the dielectric function in the vicinity of interband transitions and produces third-derivative-like spectral signatures near critical points.~\cite{Aspnes.1973} Importantly, such electroreflectance-like contributions arising from transient electric-field modulation have also been reported in pump–probe transient reflectivity studies of semiconductors.~\cite{Yang.2015,Chen.2019,Zhu.2011,Xu.2021,Pan.2024,Chen.2020,Prabhu.2004,Glinka.2008} Because PR selectively probes field-induced modulations under steady-state conditions, it provides a powerful reference for identifying analogous contributions within TA spectra. 

Here, PR measurements were performed using a chopped 405\,nm CW laser with an intensity of 1\,mW\,cm$^{-2}$ to periodically modulate the internal electric field, while the reflected light intensity of a monochromatic probe beam was recorded as a function of photon energy. 
The resulting PR spectrum, shown in \autoref{fig:res:tas}~(b), is given by the normalized change in reflectivity ($\Delta R/R$) and exhibits the characteristic third-derivative-like lineshapes associated with modulated interband transitions. The dominant features appear in the near-edge region, coinciding closely with the bleach features observed in TA. In addition, a weaker derivative-like feature is also observed in the above-edge region, near the spectral position of the strong PIA observed in TA.

To quantify the critical point energies, the PR spectrum was analyzed by fitting the data with third-derivative functional form (TDFF) functions, given by:~\cite{Aspnes.1973,Aspnes.1971,Madelung.1970,Komkov.2021}
\begin{equation*}
    \Delta R/R = \sum_j\Re[C_je^{i\theta_j}(E-E_{g,j}+i\Gamma_j)^{-n_j}] \quad ,
\end{equation*}
which describes the third-derivative-like perturbation of critical-point transitions by an electric field (low-field limit). Here, $C_j$ is the amplitude, $\theta_j$ a phase factor, $E_{g,j}$ the critical point energy, and $\Gamma_j$ the energetic broadening of the $j$-th critical point transition. The exponent $n_j$ describes the critical point type, and is given by $n=2.5$ for three-dimensional $M_0$-type interband critical points typical of bulk semiconductors, such as the \ta thin films investigated here. In addition, to account for the broad sub-gap optical response, a Gaussian component was included in the fit, as discussed below.

Analysis of the experimental PR spectrum reveals that three components are required to fit the near-edge and above-edge spectral response (see \autoref{fig:res:tas}~(b) and Table~S1). In particular, the near-edge region cannot be described by a single critical point transition, but instead comprises two components. The lower energy feature is centered at 2.14\,eV, which agrees well with the reported fundamental bandgap of \ta and coincides with the lower energy bleach observed in TA.~\cite{Eichhorn.2021,Xiao.2020,Nurlaela.2016} Thus, we attribute this feature to the fundamental band-edge transition. A second near-edge critical point is resolved at 2.27\,eV. This higher energy component exhibits a larger broadening parameter and a different phase relative to the 2.14\,eV transition, confirming that it corresponds to a separate optical resonance. 

The presence of two closely spaced near-edge features is consistent with the known optical anisotropy of \ta. Indeed, prior polarization-resolved measurements of epitaxial \ta films have revealed two absorption edges, one at 2.12\,eV for $E \parallel [100]$ and another at 2.27\,eV for $E \parallel [001]$, which correspond to inequivalent near-edge interband transitions separated by approximately 0.15\,eV.~\cite{Wang.2021} Because our films are polycrystalline, the measured PR response represents a polarization-averaged superposition of contributions from differently oriented grains. Therefore, the simultaneous observation of two closely spaced near-edge critical points with a comparable energy separation is consistent with the anisotropic band structure of \ta. Importantly, variations in thin film texture may modify the relative spectral weight of these transitions and thus influence the apparent shape of the absorption edge, potentially contributing to differences in previously reported ground state absorption and TA spectra of \ta.

Both near-edge PR components lie within the spectral region of the two bleach features observed in TA, suggesting that these transient bleach signals arise from the same critical point transitions. However, subtle differences in spectral positions are observed between the two measurements, as exemplified by the apparent $\sim100$\,meV blueshift of the higher-energy TA bleach relative to the corresponding PR critical point. These shifts can be understood by considering the fundamentally different relationships between the measured signal and the underlying dielectric function in the two measurement geometries. Measurements in reflection and transmission geometries can exhibit different apparent spectral feature positions near the same transitions, even though both are proportional to the third derivative of the dielectric function, as reflection-based measurements such as PR are more strongly influenced by the real part of the dielectric function, whereas transmission-based measurements are more strongly influenced by its imaginary part~\cite{SERAPHIN.1966}. This is confirmed by phototransmittance (PT) measurements performed under otherwise identical conditions to PR, which yield $\Delta T/T$ spectra whose features are shifted relative to PR but are closely aligned with those observed in TA (Figure~S3).

A third critical point is observed at 2.77\,eV, corresponding to an additional interband transition in the above-edge region. The energetic position of this feature is closely aligned with that of the dominant high-energy PIA observed in TA at 2.80\,eV, indicating that this transient response arises from an intrinsic critical point transition in the electronic structure of \ta. Importantly, this TA signal occurs well above the bandgap and appears as a strong PIA. Therefore, it cannot be readily explained by conventional state filling, indicating that an alternative excited-state mechanism must govern the response in this spectral region, as discussed later. 

In addition to the critical points described above, the PR spectrum exhibits a broad sub-gap signal extending toward the low-energy limit of the measurement window. Such sub-gap responses are typically associated with optical transitions involving defect states, such as \ch{O_{N}^{}} and \ch{v_{N}^{}} centers located inside the \ta bandgap.~\cite{Fu.2020,Xiao.2020} Because these localized states do not comprise critical point transitions, they are not expected to exhibit TDFF-like lineshapes and, therefore, the spectral response in this region was fitted using a Gaussian component (see \autoref{fig:res:tas}~(b) and Table~S1). Here, the sub-gap PR signal likely arises from trapping- and SPV-induced changes in defect occupancies upon illumination, consistent with the known sensitivity of \ch{O_{N}^{}} and \ch{v_{N}^{}} to the Fermi level position, as discussed further below. Having established the critical point transitions and their energetic positions, along with initial indications for the presence of optically active sub-gap defect states, we next investigate how modulation of the internal electric field manifests in the steady-state optical response of \ta using potential-dependent UV-Vis spectroscopy.

\subsection{Electric field-induced optical response}
The PR measurements above establish that the near-edge TA features coincide with electro\-reflectance-active critical points, suggesting that modulation of the internal electric field may govern the transient near-edge response. In TA, the above bandgap pump pulse generates a large density of excess charge carriers that can screen the built-in electric field near the surface and transiently flatten the bands. To independently isolate the impact of such changing internal electric fields on the optical response of \ta, we performed \puv measurements in transmission geometry using a three-electrode photoelectrochemical cell filled with 1\,M KPi (pH 12.3). Spectra were recorded as a function of the applied potential on the \ta working electrode, which was stepped from 0.0 to 1.2\,V$_\text{RHE}$. The measured potential-dependent absorption spectra were referenced to the spectrum recorded at 0.0\,V$_\text{RHE}$, which is near the flat band potential ($V_\text{fb}$) of \ta.~\cite{Fu.2022,Kawase.2022} As such, increasing the anodic bias leads to a systematic increase in the spatial extent of the depletion region at the semiconductor/electrolyte interface. 

\begin{figure}
    \centering
    \includegraphics[]{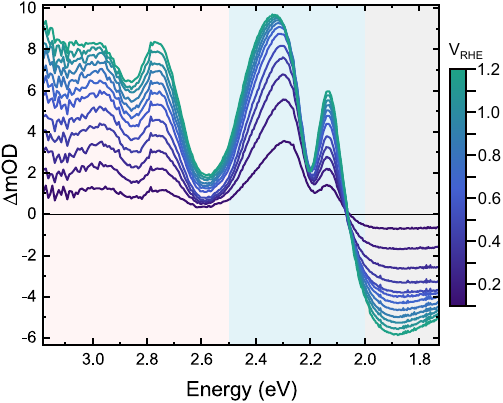} 
    \caption{Potential-dependent UV-Vis (\puv) spectra of a \ta working electrode, recorded in transmission geometry in a three-electrode electrochemical cell as a function of the applied electrochemical potential, which was swept from 0.0 to 1.2\,V$_\text{RHE}$ with a 0.1\,V step size in 1\,M pH 12.3 KPi buffer. Data are referenced to the spectrum recorded at 0.0\,V$_\text{RHE}$. The near-edge region exhibits two field-induced absorption features near 2.10\,eV and 2.35\,eV that grow with increasing anodic bias, consistent with an electroabsorption response arising from widening of the near-surface depletion region. A broad sub-gap bleach below 2.0\,eV is assigned to bias-induced change of defect charge states within the depletion region.}
    \label{fig:res:p-uv-vis}
\end{figure}

The resulting differential \puv spectra (\autoref{fig:res:p-uv-vis}) are characterized by a bleach and several field-induced absorption (FIA) features, all of which increase in absolute amplitude with increasing anodic bias. In general, the spectral response is characterized by an increasing FIA for transitions near and above the band edge and a bleach of transitions in the sub-gap range. In the above-edge region near 2.80\,eV, a weak derivative-like feature is observed, consistent with the electroreflectance response of the higher-energy critical point identified in PR. However, its relative amplitude is substantially smaller than the dominant PIA observed at this energy in TA. In the near-edge region, two FIA features are observed at approximately 2.10\,eV and 2.35\,eV, with spectral positions that are closely aligned with the critical point transitions identified by PT. Overall, the near-edge spectral response is characteristic of an electroabsorption process involving the two closely spaced interband transitions of the anisotropic band structure of \ta. 

\FloatBarrier

To further characterize the near-edge spectral evolution under an applied potential, the \puv spectra were decomposed using Gaussian functions. As shown in Figure~S4~(a), the lower-energy feature near 2.10\,eV is well described by a single component. In contrast, the higher-energy feature near 2.35\,eV requires two components to capture its asymmetric broadening toward higher photon energies with increasing anodic bias. Analysis of the potential dependence of the fitted Gaussian components reveals that the amplitudes of the two primary features at 2.15\,eV and 2.35\,eV scale approximately with $\sqrt{V}$, consistent with electroabsorption effects in the depletion region of n-type \ta, whose width increases with increasing anodic polarization as $W \propto \sqrt{V-V_\text{fb}}$.~\cite{Schmickler.2010} 
In contrast, the amplitude of the intermediate Gaussian component at 2.25\,eV initially increases but saturates at moderate anodic bias, suggesting that it originates from a different spatial region that becomes fully depleted under a moderate anodic bias of 0.4\,V$_\text{RHE}$. This behavior is consistent with the presence of a finite near-surface layer,~\cite{Fu.2022,Kawase.2022} such as amorphous layers or oxynitride phases that are known to form on the surface of \ta~\cite{He.2016,Liu.2016,Hara.2003,Li.2013,He.2019}. 

Notably, the near-edge \puv and TA spectra exhibit closely aligned spectral features, but with opposite signs. This can be understood by recognizing that the two measurements apply opposite perturbations to the internal electric field below the surface of the semiconductor. Increasing the anodic bias in \puv enhances the electric field strength within the depletion region, whereas the pump-induced SPV in TA transiently screens it. Together, these observations suggest that the near-edge TA bleach features arise from modulation of the internal electric field rather than conventional state filling or excited-state absorption processes. While such responses have previously been reported in ultrafast pump-probe studies of single crystalline III-V semiconductors measured in reflection geometry,~\cite{Yang.2015,Xu.2023,Chen.2019,Zhu.2011,Xu.2021,Pan.2024,Chen.2020,Prabhu.2004,Glinka.2008}
the present results demonstrate that analogous contributions can be resolved in transmission-mode \textmu s-s TA of polycrystalline thin film photoelectrodes. Importantly, isolating such a response enables optical tracking of photovoltage generation, charge separation, and recombination dynamics at functional semiconductor interfaces. 

In contrast to the near-edge region, where the optical response is governed by electroabsorption at the critical point transitions, the sub-gap region below approximately 2.0\,eV is characterized by a broad, spectrally flat bleach that increases in amplitude with increasing anodic bias. Prior studies have assigned sub-gap absorption in \ta to optical transitions associated with shallow substitutional O$_\text{N}$ impurities, as well as deeper v$_\text{N}$ defects and associated reduced Ta$^\text{3+}$ states, with energetic positions located 0.1\,--\,0.6\,eV below the CBM.~\cite{Fu.2020} In the present \puv measurements, increasing the anodic bias progressively depletes the near-surface region of electrons and shifts the surface Fermi level toward midgap, modifying the charge states of these defects. As the Fermi level moves toward midgap, DFT calculations predict a redistribution of defect charge states, increasing the density of \ch{O_{N}^{*}} and \ch{v_{N}^{***}} states at the cost of \ch{O_{N}^{x}} and \ch{v_{N}^{x}}/\ch{v_{N}^{*}} states.~\cite{Wang.2015,Jing.2015,Fan.2024} The resulting reduction in the occupied defect state density decreases the availability of transitions from these in-gap states to higher lying states within the conduction band (CB), consistent with the observed sub-gap bleach. However, at anodic bias potentials in excess of approximately 0.5\,V$_\text{RHE}$, a roll-over and decrease in the relative bleach strength is observed at low photon energies (near 1.8\,eV), suggesting the emergence of an additional contribution in this spectral range. Inspection of the \puv spectra collected over a broader spectral window (Figure~S5) reveals the presence of a FIA feature near 1.6\,eV that is overlaid on the broad sub-gap bleach. The energetic position of this feature is consistent with the known range of v$_\text{N}$ states within the \ta bandgap.~\cite{Fu.2020} As these deeper states become depopulated and unoccupied, transitions from the valence band and near-valence band levels into these empty in-gap states become optically allowed, giving rise to the observed FIA. Furthermore, the energetic position of this feature at approximately 1.6\,eV is consistent with the sub-gap Gaussian component identified in the PR spectrum, providing additional support for the assignment of that response to defect-mediated optical transitions involving v$_\text{N}$ states.

Taken together, the \puv data reveal two bias-dependent optical responses in \ta. In the near-edge region, the response is governed by electroabsorption at the critical point transitions, arising from modulation of the internal electric field within the depletion region. In contrast, the sub-gap response reflects bias-induced changes of in-gap defect charge states. However, neither PR nor the P-UV-Vis can account for the dominant high-energy PIA centered at 2.80\,eV observed in TA. In particular, both PR and P-UV-Vis yield relatively weak derivative-like features near this higher energy interband transition, pointing to an additional mechanism that must govern the above-edge transient response.

\subsection{Temperature-dependent optical response}
Unlike the near-edge transitions, the pronounced above-edge TA feature at 2.80\,eV appears as a strong PIA. Conventional interband state filling would be expected to reduce absorption at such a transition, producing a bleach, whereas electroreflectance effects would yield a derivative-like lineshape analogous to that observed in PR and P-UV-Vis. Therefore, the presence of a strong PIA indicates that an additional mechanism must contribute to the transient response in this spectral region. Among the possible contributions, pump-induced lattice heating is often overlooked but can significantly modify the optical spectrum through thermally induced shifting and broadening of interband transitions.

\FloatBarrier
\begin{figure*}
    \centering
    \includegraphics[width=\linewidth]{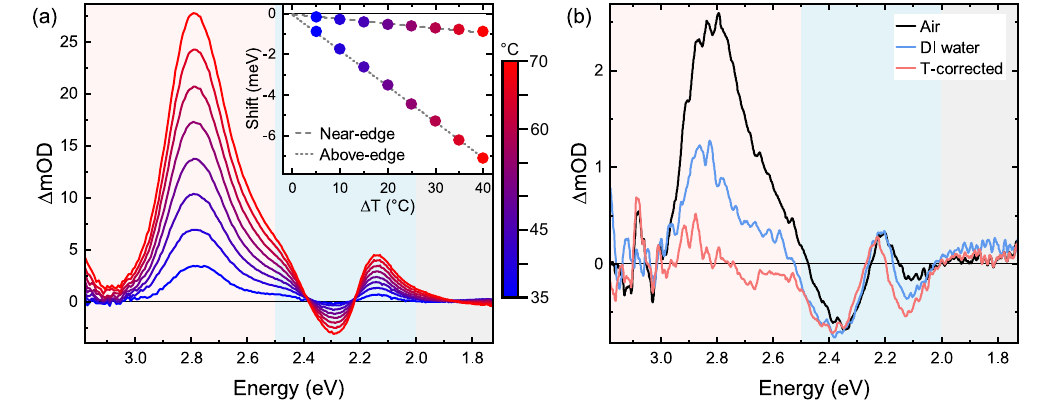} 
    \caption{(a) Temperature-dependent UV-Vis (\tuv) spectra of a \ta thin film recorded in transmission geometry as a function of temperature, stepped from 30\,$^\circ$C to 70\,$^\circ$C. Data are referenced to the spectrum recorded at 30\,$^\circ$C. A pronounced positive feature near 2.80\,eV grows with increasing temperature, closely resembling the dominant above-edge photo-induced absorption observed in TA. The inset shows the reconstructed thermally induced critical point shift for the above-edge and near-edge region. (b) Influence of pump-induced lattice heating on the TA spectrum of \ta. A TA spectrum recorded in air (black) is compared to one recorded with \ta in DI water (blue), where the higher heat capacity of water attenuates thermal contributions, confirming the thermal origin of the above-edge photo-induced absorption. Also shown is the thermally corrected TA spectrum (red), obtained by subtracting the interpolated temperature-dependent UV-Vis data from the spectrum collected in air, thereby providing the underlying electronic and defect-related contributions to the TA spectrum.}
    \label{fig:res:t-uv-vis}
\end{figure*}
To quantify the impact of lattice heating on the optical transmission of \ta thin films, we performed \tuv measurements. Transmission spectra were recorded from 30\,$^\circ$C to 70\,$^\circ$C, and were referenced to the spectrum collected at 30\,$^\circ$C (\autoref{fig:res:t-uv-vis}~(a)). The resulting differential spectra show a pronounced positive-going feature centered near 2.80\,eV, with an asymmetric tail extending toward lower photon energies, closely resembling both the lineshape and asymmetry of the dominant above-edge PIA observed in TA. This agreement strongly suggests that the above-edge PIA arises predominantly from pump-induced thermal effects rather than electronic state filling or field modulation.

In the near-edge region, the \tuv response differs markedly from the TA and PR spectra. Rather than the two closely spaced bleach features arising from the pair of interband critical points observed in those measurements, the \tuv spectra exhibit a comparatively weak, single derivative-like feature. The amplitude of this feature grows and its low-energy tail extends progressively into the bandgap with increasing temperature. Thus, although lattice heating may contribute to the near-edge TA response, thermal effects alone are insufficient to reproduce its dual-feature structure, consistent with the assignment of electronic contributions from the anisotropic band structure of \ta described above.

To quantitatively determine the extent to which lattice heating can account for the TA response, we next reconstructed the \tuv spectra from the 30\,$^\circ$C ground state absorption spectrum by incorporating energy-dependent spectral shifting and broadening parameters, as described in the Supporting Information (Figure~S6). While a single global shift and broadening parameter was not sufficient to reproduce the full spectral evolution across the measured range, this behavior is expected given the presence of multiple independent interband transitions. Using the energy-dependent parameters, the differential spectrum at each temperature could be accurately reproduced for photon energies above 2.0\,eV, as illustrated for the representative differential spectrum collected at 50\,$^\circ$C (see Figure~S6~(a)). The reconstruction confirms that the pronounced positive feature in the above-edge region is dominated by a thermally induced redshift of the higher-energy interband transition, while the weaker derivative-like structure near the absorption onset arises primarily from thermal broadening at the fundamental edge of \ta. 

The critical point energies of the principal spectral features decrease linearly with temperature, albeit with different slopes (see \autoref{fig:res:t-uv-vis}~(a) inset). Such a behavior is consistent with a Varshni-type band-gap variation over the thermal window studied here.~\cite{Varshni.1967,ODonnell.1991} The above-edge feature near 2.80\,eV exhibits the strongest temperature coefficient of -0.18\,meV\,K$^{-1}$, while the lower energy negative and positive near-edge feature varies more weakly, with a rate of -0.02\,meV\,K$^{-1}$ (see Figure~S7). This much stronger temperature dependence of the 2.80\,eV spectral feature leads to its dominant spectral weight and thermal sensitivity compared to the near-edge feature. Overall, the excellent agreement between the \tuv lineshape and the above-edge TA feature, supported by the quantitative spectral reconstruction and linear temperature scaling, indicates that pump-induced lattice heating is the primary origin of the above-edge PIA in \ta. In contrast, thermal effects alone cannot explain the dual bleach structure of the near-edge region, which is instead dominated by anisotropic electroabsorption effects, as revealed by the \puv measurements described above.

To confirm the thermal origin of the above-edge PIA, we next compared TA spectra recorded from \ta/air and \ta/DI water interfaces, exploiting the higher heat capacity of water to mitigate pump-induced heating. As shown in \autoref{fig:res:t-uv-vis}~(b), the introduction of a thin DI water layer leads to significant attenuation of the dominant above-edge PIA, further supporting the \tuv based assignment of this feature as arising from thermal effects. In contrast, much more subtle changes are observed at the lower photon energies of the near-edge and sub-gap regions. The dual-bleach structure of the near-edge region is retained upon immersion in water, but with modified relative amplitudes, as well as a slight broadening and approximately 25\,meV blueshift of the higher-energy feature. This can be understood as being a consequence of the convolution between the dominant electroabsorption response and the weaker, single derivative-like thermal contribution in this region. In particular, adding a water layer at the interface selectively reduces the thermal component while leaving the electroabsorption-derived signal mostly unchanged. 

To eliminate the thermal response and isolate the underlying electronic contributions to the TA spectrum, \tuv spectra were interpolated to match the above-edge PIA and subtracted from the TA spectrum collected in air. The resulting temperature-corrected TA spectrum is shown in \autoref{fig:res:t-uv-vis}~(b) and represents the electronic TA response following photoexcitation. Through this thermal correction procedure, independently measured \tuv data can be used to isolate the electronic and defect-related contributions to the TA spectrum, suggesting that thermal and electronic dynamics can be simultaneously tracked during pump-probe measurements.

\subsection{Thermal, electrostatic, and defect contributions to the transient optical response}

\FloatBarrier
\begin{figure}
    \centering
    \includegraphics[]{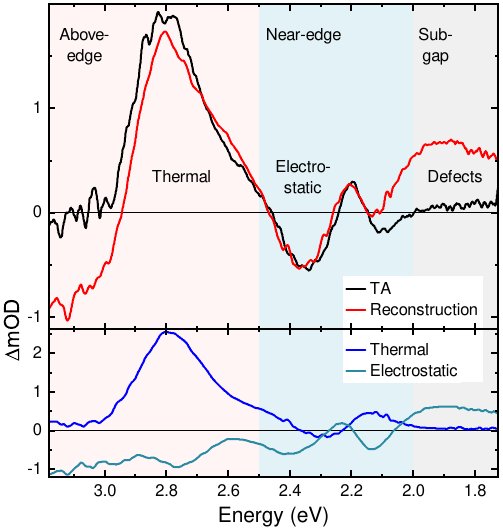} 
    \caption{TA spectrum of \ta recorded at a pump-probe delay of 1\,\textmu s (black, upper panel) together with the reconstructed spectrum obtained by co-fitting independently measured T-UV-Vis and P-UV-Vis components (red, upper panel). The individual thermal (blue) and electrostatic (teal) components are shown in the lower panel. Good agreement between the measured and reconstructed spectra is observed across the near- and above-edge spectral range, indicating that pump-induced lattice heating and transient band flattening dominate the visible TA response of \ta. Deviations between the measured and reconstructed spectra below 2.15\,eV are ascribed to contributions from non-equilibrium charge states of defects within the bandgap.}
    \label{fig:res:reconstruction}
\end{figure}

Having independently characterized the thermal and electrostatic contributions to the optical response of \ta using \tuv and \puv measurements, we now employ these data as reference spectra to decompose the TA spectrum into its constituent physical contributions. In particular, we co-fit interpolated \tuv and \puv spectra to the measured TA spectrum, initially fitting the thermal component to the above-edge region ($\geq$2.5\,eV) and the electrostatic component to the near-edge region (2.0\,--\,2.5\,eV). Due to the partial spectral overlap of these two components across the different spectral ranges, an iterative procedure was then used to achieve a fit of the experimental TA data (see Supporting Information, Section~S1). The resulting fit parameters yield an effective film temperature and degree of band flattening, allowing these physically meaningful quantities to be extracted directly from the TA data. 

\autoref{fig:res:reconstruction} shows the results of the spectral fitting procedure, with the individual thermal and electrostatic components in the lower panel and their sum compared to the experimental TA spectrum in the upper panel. Good agreement between the fitted sum and the experimental data is observed across most of the near- and above-edge spectral range. In the above-edge region ($\geq$\,2.5\,eV), the thermal component dominates and gives rise to the strong PIA centered at 2.80\,eV, consistent with pump-induced lattice heating. In the near-edge region (2.0\,--\,2.5\,eV), both thermal and electrostatic responses contribute, but with different lineshapes. The dual bleach structure in the TA spectrum is dominated by the electrostatic component, which arises from pump-induced band flattening. Together, the thermal and electrostatic components capture the near- and above-edge TA response, indicating that a vast majority of the transient optical signal in \ta across this spectral range originates from pump-induced heating and SPV modulation rather than from state filling or electronic transitions of excited carriers. The absence of state-filling contributions can be understood by considering that, at \textmu s-s pump-probe delay times, the majority of photogenerated carriers have already recombined and the remaining population primarily comprises charge-separated carriers near the semiconductor surface. At these carrier densities, band-filling effects are negligible, and the transient optical response is instead dominated by thermal and electrostatic perturbations.

The most pronounced deviation between the fitted sum and the experimental TA spectrum occurs in the sub-gap region (\textless\,2.0\,eV). Here, the electrostatic component predicts a PIA, whereas the measured TA signal is significantly weaker than predicted. While the specific origin of this discrepancy is not yet fully resolved, it likely arises from a superposition of competing PIA and bleach contributions in the sub-gap spectral range. This hypothesis is physically supported by the observation of overlapping bleach and FIA features in the \puv measurements, which arise from bias-induced changes in defect charge state populations. In this regard, it is important to recognize that \puv and TA modulate defect populations in two fundamentally different ways. In \puv, the Fermi level is shifted quasi-statically under near-equilibrium conditions, resulting in a progressive and controlled redistribution of defect charge states within the space charge region. While the SPV generated during TA leads to a similar perturbation of the built-in electric field, photoexcitation also induces a non-equilibrium splitting of quasi-Fermi levels, thereby altering both electron- and hole-active defect states. Therefore, during TA, pump-induced band flattening shifts the quasi-Fermi levels towards the band edges, giving rise to the PIA predicted from the electrostatic component, while simultaneous carrier trapping additionally modifies the occupancies and charge states of in-gap defect states, potentially giving rise to competing bleach and PIA contributions that are not captured by \puv measurements alone. For example, hole trapping at originally occupied \ch{v_{N}^{}} states would reduce the density of available \ch{v_{N}^{}} to CB transitions, leading to a sub-gap bleach~\cite{Murthy.2019}, while simultaneously opening VB to \ch{v_{N}^{}} transitions, inducing a PIA, with the net optical response depending on the relative oscillator strength of these different transitions. Furthermore, while the \puv response is confined to the space charge region, photoexcitation generates carriers throughout the full film thickness, meaning bulk defect trapping processes will also contribute to the sub-gap TA response. Overall, the weak but measurable PIA in this spectral range is indicative of a defect-related photoresponse with direct implications for the functional performance of \ta photoanodes. The complex interplay of competing contributions, as well as the deviation from simple electrostatic models revealed here, motivates future studies of these defect-mediated trapping and recombination processes.

\FloatBarrier
\begin{figure}
    \centering
    \includegraphics[]{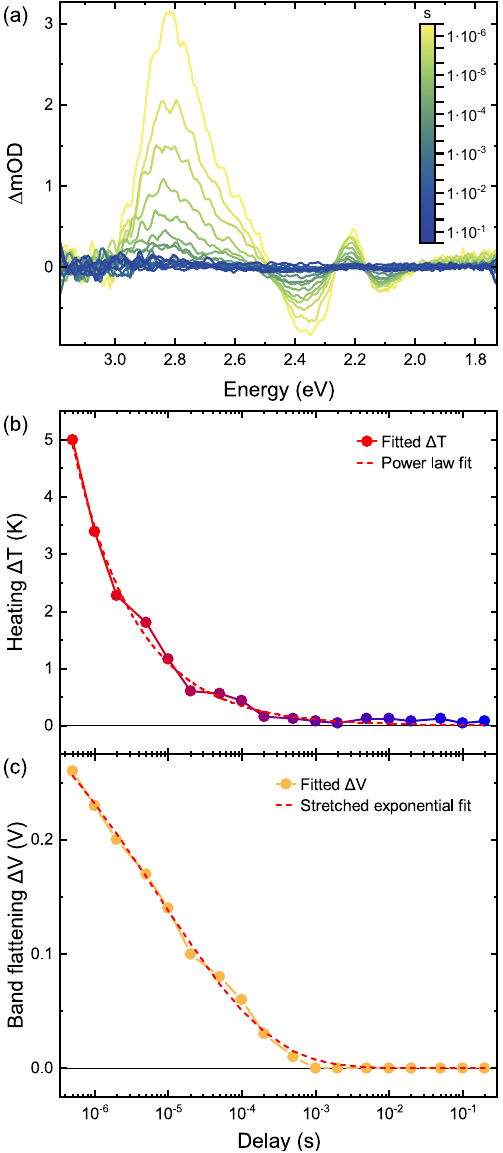} 
    \caption{(a) Full TA time delay series of a \ta thin film recorded with a pump fluence of 0.83\,mJ\,cm$^{-2}$ spanning pump-probe delays of 500\,ns to 200\,ms. (b) Transient thin film heating extracted from the co-fitting procedure as a function of pump-probe delay, together with a power-law fit (red, dashed) yielding an exponent of $b=-0.5$, consistent with one-dimensional heat diffusion into the substrate. (c) Pump-induced reduction in surface band bending extracted from the co-fitting procedure, together with a stretched-exponential fit (red, dashed) with time constant $\tau=7.93$\,\textmu s, consistent with defect-mediated recombination across an energetically distributed continuum of in-gap states.}
    \label{fig:res:processed}
\end{figure}

Having established the physical origins of the thermal, electrostatic, and defect-related contributions to the TA spectrum, we now apply the spectral decomposition approach to independently track their time evolution across the full pump-probe delay window. \autoref{fig:res:processed}~(a) shows the complete TA delay time series, spanning from 500\,ns\,--\,200\,ms, revealing a smooth decay of all features during relaxation to the ground state. By applying the fit procedure to each spectrum in the time series, we link each TA spectrum to an effective temperature and to an effective surface band flattening potential. 

\autoref{fig:res:processed}~(b) shows the extracted pump-induced lattice heating and its subsequent transient decay back to thermal equilibrium. The transient is well described by a power-law fit of the form $\Delta T = a\cdot t^b$, yielding an exponent of $b = -0.50$, in excellent agreement with the theoretically expected behavior for one-dimensional heat diffusion into the substrate.~\cite{Belliard.2015,Crank.1975} The extracted film temperature at the earliest measured delay time of 500\,ns corresponds to an increase of approximately 5\,K. This is quantitatively consistent with the upper estimate of 10.4\,K for the initial pump-induced heating (see Supporting Information, Section~S2), with the difference reflecting partial heat dissipation into the substrate prior to the first measured delay point. Thus, these results provide strong validation for the thermal assignment and deconvolution procedure, confirming that the above-edge PIA tracks the thermal transient.

\autoref{fig:res:processed}~(c) shows the extracted transient band flattening, expressed in terms of an effective applied \puv potential. This approach provides a means of tracking the SPV relaxation dynamics following photoexcitation and indicates that pump excitation induces a 250\,mV decrease in the surface band bending at the earliest measured delay time of 500\,ns. The subsequent recovery of the dark band bending follows a stretched exponential decay of the form $\Delta P = a\cdot\text{exp}(-(t/\tau)^\beta)$ with a time constant of $\tau = 7.93$\,\textmu s. Stretched exponential decay dynamics are characteristic of relaxation across an energetically distributed continuum of states rather than a single well-defined recombination pathway.~\cite{Miao.2020} This result suggests that the SPV recovery is not only governed by bulk carrier recombination but also by the trapping and detrapping dynamics of defect states within the bandgap of \ta, such as \ch{O_{N}^{}} and \ch{v_{N}^{}} discussed above, as well as surface states. As carriers are captured and released by these energetically distributed in-gap states, recombination proceeds across a range of time constants rather than through a single well-defined pathway, giving rise to the observed stretched exponential behavior. Through this analysis, we find that TA provides a powerful means of probing the transient band bending, providing additional insight into both photovoltage and recombination dynamics in \ta. Taken together, the complementary PR, \tuv, and \puv measurements indicate that the \textmu s-s TA optical response of \ta comprises three contributions that can be independently quantified -- lattice heating, surface photovoltage, and defect populations -- thereby providing key insights into the physical origins of the transient optical response of \ta.

\section{Conclusion}
In conclusion, this work demonstrates that the \textmu s-s TA spectrum of \ta comprises three spectrally distinct contributions that can be independently disentangled through a combination of PR, T-UV-Vis, and P-UV-Vis spectroscopies. Using PR, we identified three optical critical points in the band structure of \ta, including two anisotropic near-edge transitions (2.14\,eV and 2.27\,eV) and a higher-lying transition (2.80\,eV), thereby providing a basis for interpreting TA spectral features in terms of the underlying electronic structure. In the above-edge region, T-UV-Vis reveals that a dominant and previously unreported PIA centered at 2.80\,eV arises from pump-induced lattice heating, manifesting as a thermally driven redshift of the higher-energy interband transition. In contrast, a much weaker coefficient for the change of the fundamental bandgap with temperature leads to a weaker thermal component in the near-edge region. Based on these T-UV-Vis data, we report a thermal correction procedure that provides a generalizable means of removing pump-induced heating contributions from \textmu s-s TA data during post-processing, enabling access to electronic and defect-related dynamics.

Complementary P-UV-Vis measurements indicate a strong near-edge electroabsorption response arising from modulation of the surface photovoltage. The resulting pair of P-UV-Vis features is closely aligned with the dual-bleach features observed in TA, indicating that this response is dominated by transient band flattening. Fitting of the thermal and electrostatic responses to TA spectra enables quantification of a 250\,mV pump-induced reduction in surface band bending that recovers with stretched-exponential dynamics, consistent with defect-mediated recombination across an energetically distributed continuum of in-gap states. Finally, in the sub-gap region, deviations between the fitted and measured spectra suggest a complex combination of competing bleach and PIA contributions arising from both the induced surface photovoltage and the non-equilibrium redistribution of defect charge states under photoexcitation.

Notably, the absence of band-edge state-filling contributions or optical transitions of these excited-state populations stands in contrast to common interpretations of TA spectra. However, in the \textmu s-s time delay range probed in this work, the majority of photogenerated carriers have recombined and the remaining population predominantly comprises charge-separated carriers near the semiconductor surface. Such a regime is particularly important for probing competitive recombination and interfacial charge transfer associated with the chemical dynamics of functional photoelectrodes. In addition, the identification of specific spectral responses associated with lattice heating, transient photovoltages, and defect populations also provides prospects for improved interpretation of ultrafast TA spectra that include additional contributions from, for example, state-filling, bandgap renormalization, and excited-state transitions. Thus, direct comparison of the present \textmu s-s results with ultrafast fs-ns TA measurements is expected to offer a powerful opportunity to isolate and independently quantify such overlapping contributions across the full temporal range.

More broadly, the ability to independently track thermal, electrostatic, and defect-mediated contributions through rational deconvolution of TA opens new possibilities for operando characterization of photoelectrochemical and photocatalytic systems, where competitions between fast electronic relaxation and slow chemical dynamics govern energy conversion efficiencies. Moreover, this approach may be of particular relevance to the emerging field of dynamic and resonant catalysis, in which photo-activated systems exploit transiently modulated electric fields, temperatures, and charge transfer processes to enhance catalytic activity and steer selectivity.~\cite{Monai.2026} By quantifying each of these transient responses independently, the present approach represents a powerful operando tool for characterizing how external stimuli, including photoexcitation, impact physical and chemical dynamics at functional semiconductor interfaces.


\section{Experimental Section}\label{sec:details}
\subsection{Synthesis of \ta thin films} \label{sec:method:synthesis}
\ta thin films were prepared via a two-step synthesis route consisting of reactive sputter deposition of an oxide precursor (TaO$_x$) followed by high-temperature ammonolysis to form crystalline \ta. TiN/TaO$_x$ bilayer films were deposited via reactive magnetron sputtering (PVD 75, Kurt J. Lesker) onto fused silica (Siegert Wafer) substrates. As previously reported, the thin TiN underlayer serves as an electrically conducting back contact with high optical transmission across the visible range~\cite{Wagner.2026}. The sputtering system was operated at a base pressure of $5\times10^{-8}$\,Torr with a target-substrate distance of approximately 17\,cm. Before deposition, the substrates were sequentially cleaned with deionized water, acetone, and isopropanol, then dried under flowing nitrogen. For target conditioning, both the Ti (99.995\,\%, Kurt J. Lesker) and Ta (99.95\,\%, Kurt J. Lesker) targets were sputter-cleaned in an 8.5\,mTorr Ar plasma (60\,W DC, 99.9999\,\%, Linde) for 10\,min. TiN films were sputtered for 5\,min at 60\,W under 19\,sccm Ar and 1\,sccm N$_2$ flow at a substrate temperature of 600\,$^\circ$C. Subsequently, TaO$_x$ precursor films were sputter-deposited for 40\,min at the same power, temperature, and process pressure in a 10\,\% O$_2$/Ar gas mixture. During deposition, the substrate was rotated at 10\,rpm to ensure uniform film growth. Following deposition, the oxide precursor films were converted to \ta by annealing in a quartz tube furnace (Nabertherm RS 80/300/1) under a continuous NH$_3$ flow (100\,sccm, 1\,bar). The temperature was ramped at 30\,$^\circ$C\,min$^{-1}$ to 920\,$^\circ$C and held for 3\,h to promote complete nitridation and crystallization. The samples were then cooled under NH$_3$ flow until 400\,$^\circ$C, after which the clam-shell of the furnace was opened to accelerate cooling. Once at room temperature, the gas flow was switched to N$_2$ for 10\,min before the sample was removed. This procedure yielded a final film thickness of $\approx 7$\,nm for the TiN and $\approx 220$\,nm for the \ta layer.

\subsection{Structural and photoelectrochemical characterization} \label{sec:method:char}
The phase purity and morphology of the nitrided films were evaluated by GI-XRD and AFM, as previously reported for similar synthetic processes.~\cite{Wolz.2024,Wagner.2024,Eichhorn.2021} GI-XRD measurements were performed using a Rigaku SmartLab diffractometer equipped with a Cu anode and a HyPix-3000 2D detector. Data were collected at a fixed incidence angle, $\omega$, of 0.4$^\circ$ while scanning 2$\theta$ from 15$^\circ$ to 70$^\circ$ at a scan rate of 2\,$^\circ$\,min$^{-1}$ with a step size of 0.04$^\circ$. AFM images were acquired using a Bruker MultiMode system operated in tapping mode with an NSG30 cantilever (nominal tip radius 6\,nm).

PEC performance was evaluated using a home-built one-compartment three-electrode cell, with the \ta\ film serving as the working electrode, a coiled Pt wire as the counter electrode, and a low-leakage Ag/AgCl reference electrode (Driref-2SH, WorldPrecision Instruments). Potentials were controlled using a BioLogic (SP-300) potentiostat. The electrolyte consisted of 1\,M phosphate buffer (KPi, $\geq$\,98\,\%, K$_2$HPO$_4$ and K$_3$PO$_4$, Sigma-Aldrich) at pH 12.3 with 0.1\,M K$_4$[Fe(CN)$_6$] ($\geq$\,98.5\,\%, Sigma-Aldrich) added as a sacrificial hole acceptor. The ferrocyanide redox couple was employed to probe interfacial hole extraction while minimizing kinetic limitations associated with sluggish water oxidation. Prior to measurement, the electrolyte was degassed with Ar (99.9999\,\% Linde) for 10\,min. Illumination was provided by a solar simulator (Asahi, AM 1.5G) adjusted to 100\,mW\,cm$^{-2}$ at the sample position. Linear sweep voltammetry under chopped illumination was used to assess the steady-state photocurrent response, with all potentials reported with respect to the reversible hydrogen electrode (RHE).

\subsection{Transient absorption spectroscopy} \label{sec:method:tas}
For \textmu s-s TA measurements, samples were excited using a pump laser (SpitLight Compact 200 OPO, InnoLas Laser) delivering 5\,ns pulses with tunable wavelength and repetition rate under ambient conditions. For the measurements presented here, the pump wavelength was fixed at 355\,nm and the repetition rate at 1\,Hz. The pump beam was expanded to a spot diameter of approximately 1\,cm to ensure homogeneous excitation across the probed region. The pump fluence was controlled using neutral-density filters. The transient response was probed in transmission geometry using continuous wave light from a xenon lamp (ILD-Xe-QH xenon-QTH source, Bentham, 75\,W xenon lamp) directed through a grating monochromator (TMc300, Bentham) placed before the sample. For the broadband measurements presented in this work, a mirror mounted on the grating turret of the monochromator was used to deliver the full xenon lamp spectrum. Both pump and probe beams were incident on the sample from the film side, and a notch filter was placed after the sample to suppress scattered pump light. The transmitted probe light was dispersed by a second grating spectrometer (Kymera 328i, Oxford Instruments) and detected using an amplified, Peltier-cooled, gated sCMOS camera (iStar, Oxford Instruments). Data acquisition was synchronized to the pump pulse with a trigger diode (PDA8A/M, Thorlabs) placed adjacent to the sample. Unless otherwise noted, spectra were collected with a pump-probe delay of 1\,\textmu s. The resulting TA spectra are presented on an optical density scale as a function of probe beam energy $E$ and pump-probe delay time $t$:
\begin{equation*}
\Delta \mathrm{OD}(\lambda,t) = \log_{10}\left(\frac{\mathcal{T}_0(\lambda)}{\mathcal{T}(\lambda,t)}\right),
\end{equation*}
where $\mathcal{T}_0(\lambda)$ and $\mathcal{T}(\lambda,t)$ denote the probe transmission in the absence and presence of pump excitation, respectively. Spectra were averaged over 300–1800 consecutive laser pulses to improve the signal-to-noise ratio. For visualization, spectra were smoothed with a Savitzky–Golay filter (10\,nm window, fourth-order polynomial) after verification that smoothing does not alter peak positions or spectral trends. All fitting was performed prior to smoothing.

\subsection{Photoreflectance and phototransmittance spectroscopy} \label{sec:method:pr}
For PR and PT measurements, samples were excited using a 405\,nm continuous wave laser (Cobolt 06-MLD 405\,nm, Cobolt AB) operated at an intensity of 1\,mW\,cm$^{-2}$ and modulated at 333\,Hz under ambient conditions. The excitation beam was incident on the sample surface along the surface normal. Reflection and transmission changes were probed over a 1\,cm diameter active area using monochromatic light provided by the same xenon source and monochromator employed for TA measurements (see \autoref{sec:method:tas}). The monochromatic probe light was incident on the sample at an angle of 27$^\circ$ relative to the surface normal and was scanned through the spectral region of interest to construct the PR spectrum. Reflected/transmitted probe light was passed through a 405\,nm long-pass filter to suppress scattered pump light and was detected using a battery-powered Si photodiode (DET36A2, Thorlabs). The photodiode signal was fed into a transimpedance amplifier with variable gain (DLPCA-200, Femto) and was analyzed with a lock-in amplifier (Model SR830 DSP, Stanford Research Systems) referenced to the pump modulation frequency. The lock-in simultaneously recorded both the DC and AC components of the reflected/transmitted signal, enabling calculation of the normalized photoreflectance/phototransmittance response, $\Delta I / I = I_{\mathrm{AC}} / I_{\mathrm{DC}}$. 

\subsection{UV-Vis spectroscopy} \label{sec:method:uv-vis}
UV-Vis measurements were performed using a Cary 5000 UV-Vis-NIR spectrophotometer (Agilent) in transmission geometry. Spectra were acquired with a step size of 1\,nm, a spectral bandwidth of 2\,nm, and an averaging time of 1\,s per data point. Custom sample holders were used for both \tuv and \puv measurements. The \tuv cell comprised a polyether ether ketone (PEEK) housing with two quartz windows aligned along the transmission path. The sample was clamped between two copper plates with aligned apertures, enabling transmission measurements. A Pt100 thermocouple (KS-PT100-2L-1.0-330, otom Group GmbH) embedded in the Cu holder in close proximity to the sample was used to monitor the temperature, which was controlled using two Peltier elements (QC-71-1.4-8.5MS, Quick-Ohm Küpper \& Co. GmbH) mounted to the Cu structure regulated by a commercial Peltier controller (TEC-1161-10A-PT100-SCREW, Meerstetter Engineering GmbH). A custom code was used to synchronize Peltier control and UV-Vis data acquisition. The temperature was stepped from 30\,$^\circ$C to 70\,$^\circ$C. Prior to data collection at each set-point temperature, a 25\,min stabilization time was implemented to ensure thermal equilibrium. The transmission spectrum of the empty cell (without film or substrate) was used as a reference, and the absorbance $A$ at temperature $T$ was calculated as
\begin{equation*}
    A(\lambda,T) = \log_{10}\left(\frac{\mathcal{T}_{\mathrm{empty}}(\lambda)}{\mathcal{T}(\lambda,T)}\right)
\end{equation*}
where $\mathcal{T}_{\mathrm{empty}}(\lambda)$ is the transmission of the empty cell and $\mathcal{T}(\lambda,T)$ is the transmission of the mounted sample at temperature $T$. Differential \tuv spectra were then obtained relative to a reference temperature $T_0 = 30\,^\circ$C as
\begin{equation*}
    \Delta A_T(\lambda,T)
    =
    A(\lambda,T) - A(\lambda,T_0).
\end{equation*}

For \puv measurements, a custom one-compartment three-electrode photoelectrochemical cell with a PEEK body and fused silica window was used. The substrate served as the other optical window with a 7\,mm diameter opening in the housing allowing for transmission measurements under applied bias. The electrode configuration was identical to that used for PEC characterization, except that the sacrificial hole scavenger was omitted from the electrolyte. Potentials were applied in chronoamperometric mode with a BioLogic (SP-300) potentiostat. A custom code was used to synchronize the application of potential steps and acquisition of UV-Vis spectra. To equilibrate the sample, it was held at 0\,V$_\text{RHE}$ for 4\,h in the dark before starting the UV-Vis measurements. In addition, prior to each acquisition, the sample was allowed to equilibrate at the set-point potential for 5\,min. Transmission spectra were referenced to a bare quartz substrate and the absorption spectra at applied potential $V$ were calculated as 
\begin{equation*}
    A(\lambda,V) = \log_{10}\left(\frac{\mathcal{T}_{\mathrm{sub}}(\lambda)}{\mathcal{T}(\lambda,V)}\right)
\end{equation*}
where $\mathcal{T}_{\mathrm{sub}}(\lambda)$ is the transmission of the bare substrate and $\mathcal{T}(\lambda,V)$ is the transmission of the mounted sample at applied potential $V$. Differential \puv\ spectra were then obtained relative to a reference potential $V_0 = 0$\,V$_\text{RHE}$ as
\begin{equation*}
    \Delta A_V(\lambda,V) = A(\lambda,V) - A(\lambda,V_0).
\end{equation*}

\section{Data Availability Statement}
The data that support the findings of this study are available from the corresponding author upon request.

\section*{Acknowledgment}
This project has received funding from the European Research Council (ERC) under the European Union’s Horizon 2020 research and innovation program (grant agreement no. 864234), from the Deutsche Forschungsgemeinschaft (DFG, German Research Foundation) under Germany´s Excellence Strategy – EXC 2089/2 – 390776260, and TUM.Solar in the context of the Bavarian Collaborative Research Project Solar Technologies Go Hybrid (SolTech). This work further acknowledges funding by the Deutsche Forschungsgemeinschaft (DFG, German Research Foundation) – Grant No. 428591260.

\bibliographystyle{rsc}
\bibliography{References_abbreviated}

\end{document}


\begin{center}
    {\large \textbf{Supporting Information:}} \\[1ex] 
\end{center}    
    {\large \textbf{Quantifying Thermal, Photovoltage, and Defect Contributions to Transient Absorption of \ta Photoanodes}}  

\author{Johannes Dittloff}
\affiliation{Walter Schottky Institute, Technical University of Munich, 85748 Garching, Germany}
\affiliation{Physics Department, TUM School of Natural Sciences, Technical University of Munich, 85748 Garching, Germany}

\author{Lukas M. Wolz}
\affiliation{Physics Department, TUM School of Natural Sciences, Technical University of Munich, 85748 Garching, Germany}

\author{Matthias U. Quintern}
\affiliation{Walter Schottky Institute, Technical University of Munich, 85748 Garching, Germany}
\affiliation{Physics Department, TUM School of Natural Sciences, Technical University of Munich, 85748 Garching, Germany}

\author{Laura I. Wagner}
\affiliation{Walter Schottky Institute, Technical University of Munich, 85748 Garching, Germany}
\affiliation{Physics Department, TUM School of Natural Sciences, Technical University of Munich, 85748 Garching, Germany}

\author{Matthias Kuhl}
\affiliation{Physics Department, TUM School of Natural Sciences, Technical University of Munich, 85748 Garching, Germany}

\author{Johanna Eichhorn}
\affiliation{Physics Department, TUM School of Natural Sciences, Technical University of Munich, 85748 Garching, Germany}

\author{Ian D. Sharp$^*$}
\email[sharp@wsi.tum.de]{}
\affiliation{Walter Schottky Institute, Technical University of Munich, 85748 Garching, Germany}
\affiliation{Physics Department, TUM School of Natural Sciences, Technical University of Munich, 85748 Garching, Germany}
\pacs{}
\maketitle

\FloatBarrier
\begin{figure*}[]
    \centering 
    \includegraphics[width=\linewidth]{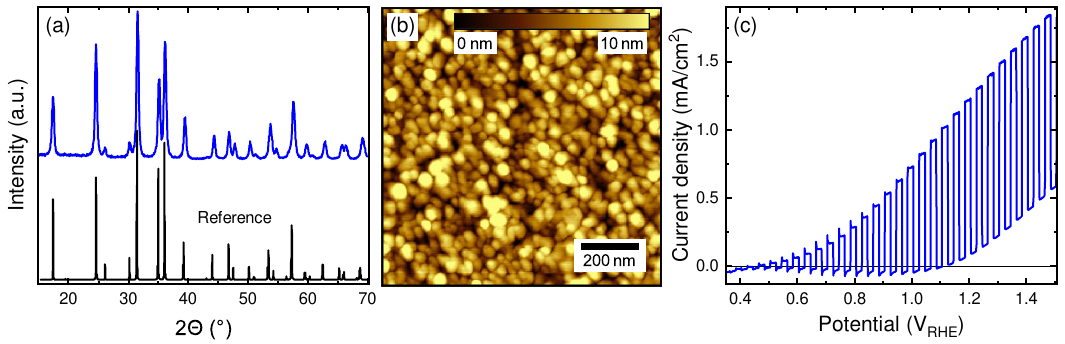}
    \caption{Structural and photoelectrochemical characterization of \ta thin films. (a) Grazing incidence X-ray diffraction pattern confirming formation of phase-pure orthorhombic \ta. (b) Atomic force microscopy image revealing a granular morphology with a root-mean-square roughness of approximately 2.4\,nm. (c) Photoelectrochemical linear sweep voltammogram measured under chopped AM1.5G illumination in 1\,M KPi buffer (pH 12.3) with 0.1\,M \ch{K_4[Fe(CN)_6]} as sacrificial hole acceptor.}
    \label{fig:si:xrd}
\end{figure*}

\begin{figure*}[]
    \centering 
    \includegraphics[]{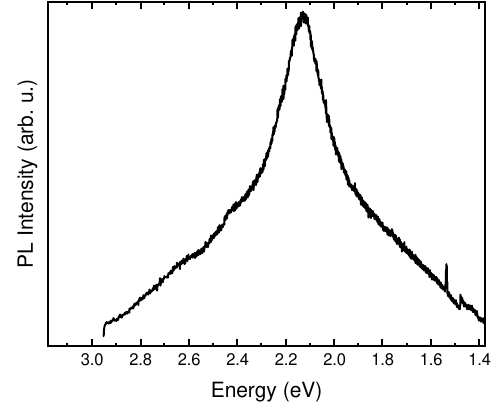}
    \caption{Room temperature photoluminescence emission spectrum recorded under 405\,nm continuous wave laser excitation.}
    \label{fig:si:pl}
\end{figure*}

\FloatBarrier
\begin{table}[]
\setlength{\tabcolsep}{12pt} 
\renewcommand{\arraystretch}{1.3}
\centering
\caption{Summary of the fitted parameters for the three-component third-derivative functional form (TDFF) lineshape fit together with a Gaussian contribution for the sub-gap range of the PR data. The following fit function was used: \newline$\Delta R/R = \sum_{j=1}^3\Re[C_je^{i\theta_j}(E-E_{g,j}+i\Gamma_j)^{-n_j}] + A\,e^{-(E-E_0)^2/2\sigma^2}$.}
\begin{tabular}{|c|c|c|c|c|c|}
\hline
TDFF component ($j$) & $C$ & $\theta$ ($^\circ$) & $E_g$ (eV) & $\Gamma$ (eV) & $n$\\
\hline
1 & 6.70$\times10^{-9}$ & 117.20 & 2.77 & 0.10 & 2.5\\
2 & 8.68$\times10^{-8}$ & 0.67 & 2.27 & 0.15 & 2.5\\
3 & 2.52$\times10^{-8}$ & 111.62 & 2.14 & 0.08 & 2.5\\
\hline
\noalign{\vskip 0.5cm} 
\hline
Gaussian & $A$ & & $E_0$ (eV) & $\sigma$ (eV) &\\
\hline
& -2.64$\times10^{-6}$ & & 1.72 & 0.16 & \\
\hline
\end{tabular}
\label{tab:si:pr}
\end{table}

\FloatBarrier
\begin{figure*}[]
    \centering 
    \includegraphics[]{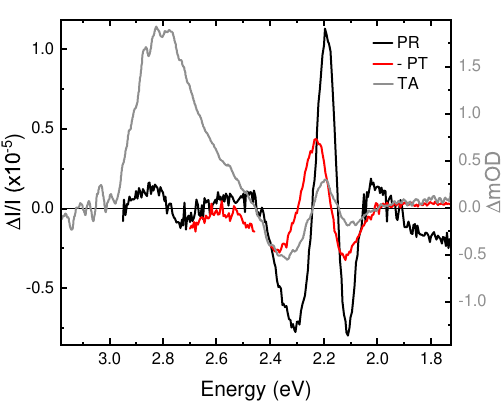}
    \caption{Comparison of photoreflectance (PR), phototransmittance (PT), and transient absorption (TA) spectra of \ta. PR and PT are plotted on the left axis and TA on the right axis. The sign of the $\Delta T/T$ PT signal is inverted to enable direct comparison with the TA data, which are presented on an absorption scale. The close alignment of PT and TA 
    spectral features confirms that the near-edge bleach positions in TA correspond to the critical point transition energies identified by PR. Spectral shifts between PR and PT/TA arise from a different sampling of light-induced changes to the dielectric function in reflection and transmission geometries, respectively.}
    \label{fig:si:pt}
\end{figure*}

\FloatBarrier
\begin{figure*}
    \centering
    \includegraphics[width=\linewidth]{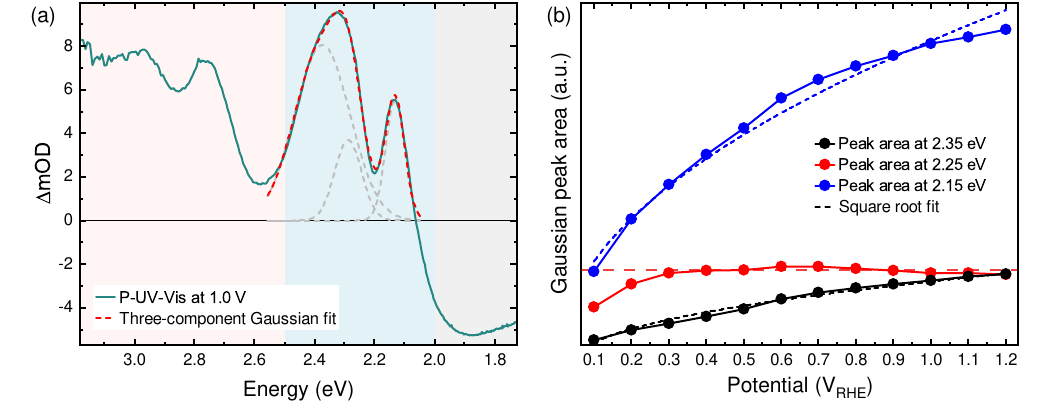} 
    \caption{Gaussian decomposition of near-edge potential-dependent absorption (\puv) features. (a) Representative \puv spectrum recorded at 1.0\,V$_\text{RHE}$, fitted with three Gaussian components at 2.15\,eV, 2.25\,eV, and 2.35\,eV (grey dashed lines). (b) Potential dependence of the fitted Gaussian areas. The areas of the 2.15\,eV and 2.35\,eV components scale approximately with $\sqrt{V - V_\text{fb}}$ (dashed black and blue lines, respectively), consistent with electroabsorption in the depletion region of an n-type semiconductor. In contrast, the 2.25\,eV component saturates at moderate anodic bias (red dashed line), consistent with full depletion of a finite near-surface layer.}
    \label{fig:si:p-uv-vis-areas}
\end{figure*}

\FloatBarrier
\begin{figure*}
    \centering 
    \includegraphics[]{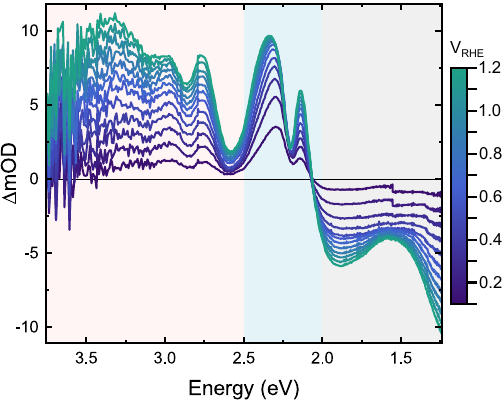}
    \caption{\puv data from the main text shown over the full recorded spectral range. The extended range reveals the emergence of an additional feature at 1.6\,eV with increasingly large anodic bias potentials.}
    \label{fig:si:p-uv-vis-full}
\end{figure*}

\FloatBarrier
\begin{figure*}
    \centering
    \includegraphics[width=\linewidth]{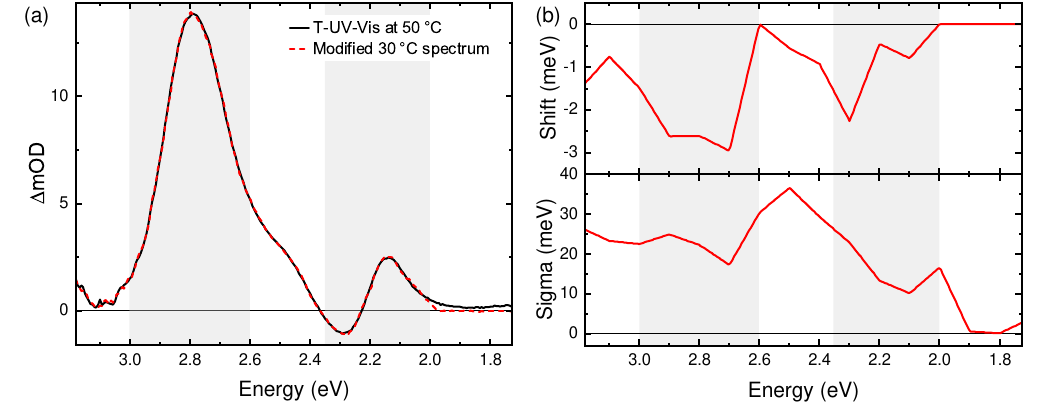} 
    \caption{Reconstruction of temperature-dependent UV-Vis (\tuv) spectra by applying energy-dependent spectral shift and broadening parameters to the 30\,$^\circ$C reference spectrum. (a) Comparison of the experimental (black) and simulated (dashed red) differential \tuv spectrum at 50\,$^\circ$C. Good agreement between the experimental and simulated spectra confirms that the \tuv response above 2.0\,eV can be accurately described by thermally induced shifts and broadening of interband transitions. (b) Energy-dependent shift (top) and broadening (bottom) parameters used in the simulation.}
    \label{fig:si:t-uv-vis-continous}
\end{figure*}

\FloatBarrier
\begin{figure*}
    \centering
    \includegraphics[width=\linewidth]{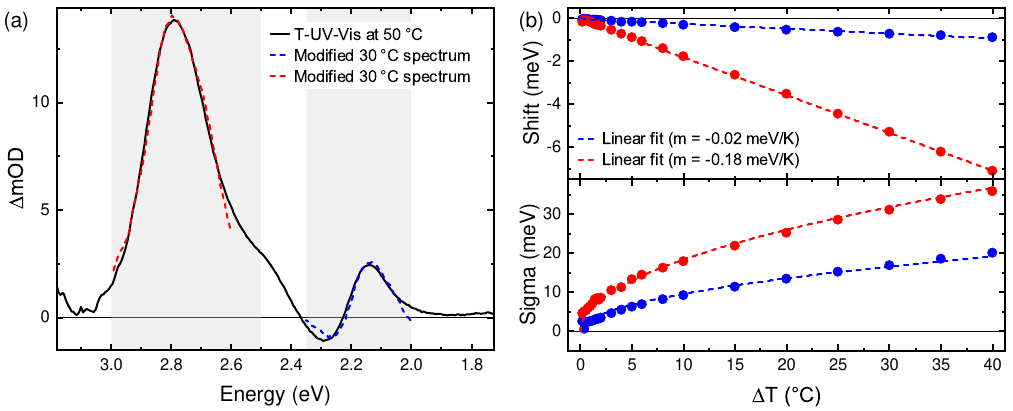} 
    \caption{Reconstruction of the main features of \tuv spectra using a non-continuous approach compared to \autoref{fig:si:t-uv-vis-continous}. Two energy windows (2.50\,--\,3.00\,eV and 2.00\,--\,2.35\,eV) containing the main features are used for reconstruction. In each window only one set of shift and broadening parameters is utilized. (a) Exemplary comparison of the experimental (black) and piecewise simulated (dashed) differential \tuv spectra at 50\,$^\circ$C. (b) Extracted shift of the critical point energy for all recorded temperatures during \tuv together with linear fits (top). Corresponding broadening parameters (bottom). The broadening scales approximately with $\sqrt{T-T_0}$.}
    \label{fig:si:t-uv-vis-multiroi}
\end{figure*}

\FloatBarrier
\begin{figure*}
    \centering
    \includegraphics[]{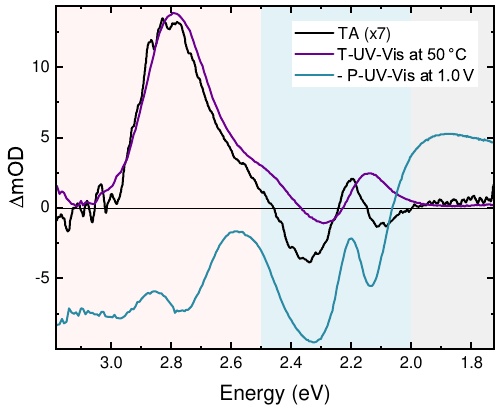} 
    \caption{Spectral comparison of the TA spectrum recorded at a pump-probe delay of 1\,\textmu s (scaled by a factor of 7 for visibility), a \tuv spectrum at 50\,$^\circ$C referenced to 30\,$^\circ$C, and a \puv spectrum at 1.0\,V$_\text{RHE}$ with inverted amplitude. The close alignment of spectral features across all three measurements demonstrates the correspondence between the transient optical response and the independently measured thermal and electrostatic modulations.}
    \label{fig:si:uv-vis-comp}
\end{figure*}

\clearpage
\section{Fitting of TA spectra}\label{sec:si:fit-spectra}
To extract quantitative information on pump-induced lattice heating and surface photovoltage from TA spectra, a Python-based iterative fitting procedure was developed. As inputs, the procedure requires a TA spectrum or time series of spectra, together with \tuv and \puv datasets. The fitting assumes that the 2.5\,--\,3.0\,eV region of the TA spectrum is dominated by thermal effects, whereas the 2.0\,--\,2.5\,eV region is dominated by band bending effects, consistent with the assignments reported in the main text. Linear interpolation of the UV-Vis datasets was applied to increase the effective temperature and potential resolution beyond the experimentally sampled values.

As shown in Figure~4 of the main manuscript, the nominally thermally and electrostatically dominated regions contain partial contributions from the respective other effect. Therefore, an iterative fitting procedure was employed. In the first iteration, interpolated \tuv data were fitted to the TA spectrum in the above-edge region, and the best-matched thermal component was subtracted from the TA spectrum. An analogous \puv fit was then performed on the thermally corrected TA spectrum in the near-edge region. The resulting electrostatic component was subtracted from the initial TA spectrum to yield a band-bending-corrected spectrum, on which a refined \tuv fit was performed. Finally, the improved thermal component was then again subtracted from the initial TA spectrum to achieve a refined \puv fit.

Several physical constraints were imposed on the fitting process. The thermal component was restricted to positive temperature changes, consistent with pump-induced lattice heating. The electrostatic component was restricted to represent a 
pump-induced reduction in near-surface band bending, consistent with surface 
photovoltage generation. Importantly, the \tuv reference temperature and \puv reference potential were treated as free parameters rather than being fixed to 30\,$^\circ$C or 0\,V$_\text{RHE}$, allowing the fit to select the optimal dark and pumped reference spectra from the full UV-Vis dataset (see Experimental Section of the main manuscript). Finally, for fits of a complete TA delay time series, an additional global constraint was applied to the \puv component, requiring that the band bending in the unpumped reference state remains constant across all pump-probe delays. The fitting procedure thus yields values for the pump-induced temperature rise and reduction in surface band bending at each delay time, enabling independent tracking of thermal and photovoltage dynamics across the complete temporal range.

\section{Pump-induced temperature change} \label{sec:si:temp}
An upper bound for the pump-induced heating of the \ta thin film during TA measurements was estimated by assuming complete absorption of a single 5\,ns pulse with a fluence of $E_\text{TA} = 1.0$\,mJ\,cm$^{-2}$ at 355\,nm within the 220\,nm thick film. This assumption is justified by the large absorption coefficient of \ta at this wavelength, $\alpha \approx 5\times10^5$\,cm$^{-1}$,~\cite{Ziani.2015} which gives complete absorption within the film thickness via the Beer-Lambert law. Since no experimental data for the specific heat capacity of \ta are available, a value of $c = 450$\,J\,kg$^{-1}$\,K$^{-1}$ was assumed, together with a mass density of $\rho = 9.85$\,g\,cm$^{-3}$.~\cite{Brauer.1966} The initial pump-induced temperature rise is then approximated as
\begin{equation*}
    \Delta T_\text{TA} = \frac{E_\text{TA}}{\rho \cdot d \cdot A \cdot c} 
    = 10.4\,\text{K}\,,
\end{equation*}
where $d = 220$\,nm is the film thickness and $A = \pi r^2$ is the pump beam area. This represents an upper bound, as heat dissipation into the substrate and ambient air, as well as pump reflection losses, are neglected.

An analogous estimate was performed for the PR measurement, in which a 405\,nm continuous wave pump at $I = 1$\,mW\,cm$^{-2}$ is modulated at $f = 333$\,Hz with a 50\,\% duty cycle. The energy deposited per modulation period is
\begin{equation*}
    E_\text{PR} = \frac{I}{2f} = 1.5\,\text{\textmu J\,cm}^{-2}\,.
\end{equation*}
Scaling by the ratio of deposited energies yields an upper bound for the PR-induced temperature rise of
\begin{equation*}
    \Delta T_\text{PR} \approx \frac{E_\text{PR}}{E_\text{TA}} 
    \cdot \Delta T_\text{TA} = 0.015\,\text{K}\,.
\end{equation*}
This value is negligible compared to the pump-induced heating in TA, confirming that lattice heating does not contribute meaningfully to the PR response and that the PR measurements can be interpreted purely in terms of electronic modulation of the optical transitions.

\bibliographystyle{rsc}
\bibliography{References_abbreviated}